\documentstyle{mn}
\def\etal{{\it et al. }}
\def\deg{\ifmmode^\circ\else$^\circ$\fi}
\def\ltsima{$\; \buildrel < \over \sim \;$}
\def\simlt{\lower.5ex\hbox{\ltsima}}
\def\gtsima{$\; \buildrel > \over \sim \;$}
\def\simgt{\lower.5ex\hbox{\gtsima}}

\newcounter{parentequation}

\begin{document}

\title[Constraints on cosmological Parameters]{
Constraints on $\Omega_\Lambda$ and $\Omega_m$ from
Distant Type 1a \\
Supernovae and Cosmic Microwave Background Anisotropies}

\author[G. Efstathiou et al.]{G. Efstathiou$^{1}$, S. L. Bridle$^2$,
A.N. Lasenby$^2$, M.P. Hobson$^2$,  R.S. Ellis$^1$\\
$^1$ Institute of Astronomy, Madingley Road, Cambridge, CB3 OHA.\\
$^2$ Cavendish Astrophysics Group, Cavendish Laboratory,
Madingley Road, Cambridge CB3 OHE.}

\maketitle

\begin{abstract}
We perform a combined likelihood analysis of the latest cosmic
microwave background  anisotropy data and distant Type 1a Supernova
data of Perlmutter \etal (1998a). Our analysis is restricted to
cosmological models where structure forms from  adiabatic
initial fluctuations characterised by a power-law spectrum with
negligible tensor component. Marginalizing over other parameters, our
best fit solution gives $\Omega_m = 0.25^{+0.18}_{-0.12}$ 
and $\Omega_\Lambda = 0.63^{+0.17}_{-0.23}$ ($95 \%$ confidence
errors) for the cosmic densities contributed by matter and a cosmological
constant respectively.  The results therefore strongly favour a nearly
spatially flat Universe with a non-zero cosmological constant.
\end{abstract}

\section{Introduction}\label{sec:intro}

The aim of this paper is to constrain the geometry of the Universe by
combining results from the cosmic microwave background anisotropies
(CMB) with those from distant Type 1a supernovae. Recently, there have
been a number of analyses of CMB anisotropies aimed at constraining
cosmological parameters (Hancock \etal 1998, Bond and Jaffe 1997,
Lineweaver and Barbosa 1998a,b, Webster \etal 1998). These papers show
that the CMB already provides useful constraints on adiabatic
inflationary models. The results described in this paper extend these
analyses to a wider parameter set including closed Universes.

 However, observations of CMB anisotropies alone cannot determine the
geometry of the Universe unambiguously (Bond, Efstathiou and Tegmark
1997, Zaldariaga, Spergel and Seljak 1997). This is because two models
with identical fluctuation spectra and matter content will have nearly
identical CMB power spectra if they have the same angular diameter
distance to the last scattering surface. This {\it geometrical
degeneracy} can be broken for extreme values of the cosmological
parameters $\Omega_\Lambda$ and $\Omega_m$ by the inhomogeneous
Sachs-Wolfe effect at low multipoles (see Efstathiou and Bond 1998),
but for plausible parameters the degeneracy is nearly exact.
\footnote{Gravitational lensing of the CMB can break this degeneracy
however (see Seljak and Zaldarriaga 1998).}

A number of authors (White 1998, Tegmark \etal 1998, Efstathiou and
Bond 1998) have shown that the magnitude-redshift relation for distant
Type 1a supernovae (SN) provides nearly orthogonal constraints in the
$\Omega_\Lambda$--$\Omega_m$ plane to those derived from the CMB.  The
combination of CMB and SN data can thus provide tight constraints on
the geometry of the Universe. This has been demonstrated by White
(1998), Garnavich \etal (1998) and Lineweaver (1998) using the SN data
of Perlmutter \etal (1997, 1998b) and Riess \etal (1998). In this
paper, we use the larger sample of $42$ high-redshift SN of Perlmutter
\etal (1998a, hereafter P98,  1998b). The likelihood analysis of the CMB
observations is described in Section 2.  Section 3 describes the
likelihood analysis of the SN data and the results from the combined
data set are described in Sections 4 and 5.

\section{Anisotropies of the Cosmic Microwave Background}

The analysis presented here is similar to that described by Hancock
\etal (1998) and we
refer the reader to this paper for technical details. Similar
analyses and compilations of observations are discussed by Lineweaver
(1998, and references therein) and by Bond, Jaffe and Knox (1998).
A window function  $W_l$ for each experiment is used to
convert the observed level of anisotropy 
to flat bandpower estimates $(\Delta T_l/T) \pm \sigma$ centred on the
effective multipole  $l_{\rm eff}$ (defined as the half-power point
of the  window function). The resulting CMB data points are plotted in
Figure 1, together with their 68 per cent confidence limits.
These confidence limits have been obtained using likelihood analyses
and hence incorporate uncertainties due to random errors, sampling
variance  and cosmic variance.

The data points  in Figure 1 are identical to those given
in Webster \etal (1998) except that we have added
the three QMAP points (Ka and Q bands for flights
1 and 2, Devlin \etal 1998, Herbig \etal 1998, 
De Oliveira-Costa \etal 1998).

\begin{figure}

\vskip 2.65 truein

\includegraphics{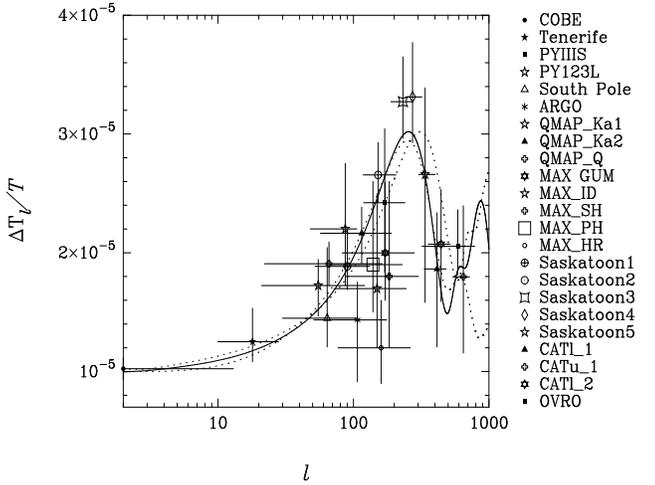}

\caption
{CMB bandpower anisotropy estimates for various experiments,
as described in the text. The solid line
shows the best fitting adiabatic CDM model with
parameters $Q_{10}=1.14$, $n=1.08$, $\omega_c = 0.36$, 
$\omega_b = 0.03$ and Doppler peak location parameter
$\gamma_D = 1.18$.  The dotted lines show the
best fitting  curves with location parameter fixed at
$\gamma_D=0.8$ and $\gamma_D = 1.5$ (approximately the
$2\sigma$  allowed range for $\gamma_D$.}

\label{figure1}
\end{figure}

We fit the CMB data points to adiabatic CDM models specified by the
following parameters: (1) amplitude $Q_{10}$ of the $\ell=10$
multipole defined as in Lineweaver (1998);
%$C_{10} = Q_{10}^2 C_{10}^{\rm COBE}$ assuming a scale-invariant
%spectrum for COBE with a  normalization of
%$Q^{RMS-PS} = 18.4 \mu K$ ({\it e.g.} Tegmark 1997); 
(2) the density parameters $\Omega_k$ and
(3) $\Omega_\Lambda$; (4) the scalar spectral index $n_s$; (5)
the physical density in cold dark matter $\omega_c = \Omega_c h^2$
\footnote{Here $h$ is the Hubble constant in units of 
$100{\rm km}{\rm s}^{-1} {\rm Mpc}^{-1}$.}; (6) the physical density
in baryons $\omega_b = \Omega_b h^2$. We ignore tensor modes
in this analysis.\footnote{For reasonable amplitudes of a tensor mode, the
effect on the position of the first Doppler peak is small. A small
tensor mode will therefore have  little effect on the cosmological
parameters $\Omega_\Lambda$ and $\Omega_m$ since these are determined
primarily by the position of the first Doppler peak.}

The motivation for choosing this set of variables is explained 
in Efstathiou and Bond (1998, hereafter EB98). 
The physical densities $\omega_c$
and $\omega_b$ and the radiation density  
determine the sound speed at the time of recombination. The
geometry of the universe is specified by the parameters $\Omega_k$
and $\Omega_\Lambda$, and the Hubble constant enters as an auxiliary
parameter 
\begin{equation}
h = \left [ {(\omega_c + \omega_b) \over 
( 1 - \Omega_k - \Omega_\Lambda) } \right ]^{1/2}. 
\end{equation}

\begin{figure}

\vskip 2.6 truein

\includegraphics{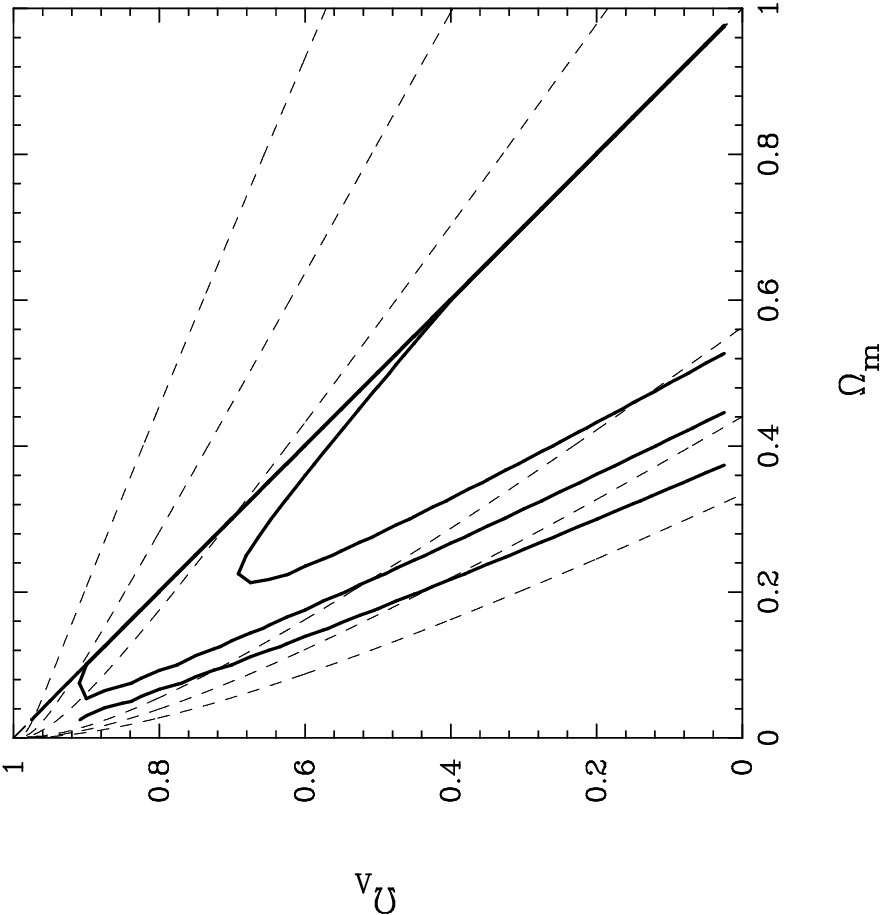}

\caption
{Likelihood contours  in the $\Omega_\Lambda$
-- $\Omega_m$ plane derived from the CMB data points shown in Figure
1. The contours are plotted where $-2{\rm ln}{\cal L}/{\cal L}_{max}$ 
is equal to $2.29$, $6.16$ and $11.83$, corresponding approximately
to $1$, $2$ and $3 \sigma$ confidence contours for a Gaussian likelihood
function.
The solid lines show the marginalized likelihood contours derived
from the CMBFAST computations for $\omega_b = 0.019$. The dotted lines 
extending into the $\Omega_k < 0$ region show the equivalent
contours derived from the fitting function approach described in the
text. For the dotted contours, we have marginalized over $\omega_b$, although
this has very little effect.}

\vskip -0.2 truein

\label{figure2}
\end{figure}

We have fitted the observations to the theoretical models using two
methods. Firstly, we compute a grid of theoretical power spectra
in the parameters $\Omega_k$, $\Omega_\Lambda$, $n_s$, $\omega_c$,
using the CMBFAST code of Seljak and Zaldarriaga (1996) with
$\omega_b$ constrained to $0.019$, the value inferred from primordial
nucleosynthesis and the deuterium abundances measured from quasar spectra
($\omega_b = 0.019 \pm 0.001$, see Burles and Tytler 1998a,b). In its 
present form, the CMBFAST code is restricted to open and spatially 
flat models ($\Omega_k \ge 0$)
 and so we have adopted a second, approximate, 
technique to extend the theoretical predictions to
closed models. This is based on a semi-analytic fitting formula
for the CMB power spectrum of $\Omega_k = 0$ models, which is 
a generalization of the fitting formula
of equation (24) in EB98 and  provides
accurate fits to the first three Doppler peaks
of the CMB power spectrum. The fitting formula includes the
dependences on $n_s$, $\omega_b$, $\omega_c$ and $\Omega_\Lambda$
and is typically accurate to better than 5 percent. For models with non-zero
values of $\Omega_k$, we use the scaling relation $C(\ell^\prime)
\rightarrow C(\ell \gamma_D)$, where $\gamma_D$ is a `location'
parameter,
\begin{equation}
\gamma_D \approx {\ell_D(\Omega_k, \Omega_\Lambda)
\over \ell_D(\Omega_k=0, \Omega_\Lambda=0)},
\end{equation}
where $\ell_D(\Omega_k, \Omega_\Lambda)$ is the location of the first
Doppler peak given by equation (22) of EB98 generalized to models with
$\Omega_k < 0$.  The location parameter thus measures the positions of
the Doppler peaks relative to those of a spatially flat model with
zero cosmological constant. The approximate formula does not include
the inhomogeneous Sachs-Wolfe effect (see {\it e.g.} Bond 1996) which
affects low multipoles if $\Omega_k$ and $\Omega_\Lambda$ are
non-zero. However, the inhomogeneous Sachs-Wolfe effect is a weak
discriminator of cosmological models (see EB98) and,  with the data
shown in Figure 1, the constraints on cosmological models are set
primarily by the location of the first Doppler peak. A similar
approximate technique, using rescaling of power spectra computed
with CMBFAST, is described by Tegmark (1998).

Figure 1 shows the best fitting CMB power spectrum, together with the
best fitting curves with $\gamma_D=0.8$ and $1.5$ (spanning the $2\sigma$
allowed range of $\gamma_D$ after marginalization over other parameters).
The present CMB data points evidently favour models with $\Omega_k \approx 0$.
The large number of data points at $\ell \sim 100$ set quite 
strong constraints on closed models, but the lack of data
points at $\ell \simgt 300$ leads to  weaker constraints on open 
models. The CMB data do not yet allow strong 
constraints on the parameters $\omega_b$ and $\omega_c$, hence the
constraints on Hubble constant (equation 1) are also extremely weak.

Figure 2 shows the CMB constraints in the $\Omega_m-\Omega_\Lambda$
plane.  Here, we have marginalized over $n_s$, $Q_{10}$, $\omega_c$ (and
$\omega_b$ for the likelihoods computed from the fitting formula)
assuming uniform prior distributions in these parameters. The
marginalized likelihood depends slightly on the range adopted for
$\omega_c$; in Figure 2 we assume a uniform prior distribution over
the range $0.05 \le \omega_c \le 0.5$.  The marginalized likelihood
function is insensitive to the ranges and priors adopted for the other
parameters and is insensitive to $\omega_b$. 

The results of this analysis are similar to those of Lineweaver (1998)
and Tegmark (1998) but differ in detail. The main difference is in the
way that we marginalize over the likelihood function. We have assumed
a uniform prior distribution in each parameter and performed direct
integrations over the full likelihood function. Lineweaver and Tegmark
`marginalize' over a subset of parameters by fixing them to their
maximum likelihood values. Our approach provides more robust errors,
though the marginalized likelihood function for poorly determined
parameters will depend on the choice of prior (usually weakly).  In
the limit that the likelihood function is Gaussian, the two approaches
are equivalent. However, for non-Gaussian likelihoods (as is the case
with the present CMB data) the  approach adopted by 
Lineweaver and Tegmark can give
misleadingly small errors on some parameters.

\section{Magnitude-Redshift Relation for Distant Supernovae}

We use the sample of $42$ high redshift ($0.18 \le z \le 0.83$)
supernovae of P98, supplemented with $18$ low redshift ($z < 0.1$)
Type 1a supernovae from the Cal\'an/Tololo Supernova Survey (Hamuy
\etal 1996). For each supernova, P98 computed a peak magnitude in the
B band $m_B$, corrected for Galactic extinction  and a `stretch 
parameter' $s$ that stretches the time axis of
a template Type 1a lightcurve to match the observed lightcurve (see
Perlmutter \etal 1995, 1997).

P98 provide a comprehensive analysis of the constraints on $\Omega_m$
and $\Omega_\Lambda$ derived from the SN magnitude-redshift relation
and of the effects of excluding
various outlying SN,   including or excluding corrections for the
lightcurve width-luminosity relation,  host galaxy
extinction, {\it etc}. P98 show
that the likelihood function in the $\Omega_m$--$\Omega_\Lambda$ 
plane is remarkably stable to such changes. We do not repeat this analysis
here, but instead concentrate on the analysis of the supernovae used in
the `primary fit' of P98 (their fit C) which excludes 4 high redshift
objects.  These are,
SN 1997O, 1996cg and 1996cn, which are very likely reddened by their
host galaxies and so are fainter than the best fitting magnitude
redshift relation,  and SN 1994H which is not spectroscopically
confirmed as a Type 1a SN and lies brighter than the best-fit
relation. In agreement with the results of P98, none of our conclusions
change significantly if we include these supernovae.

We define a corrected peak magnitude $m_B^{\rm corr}$ for the lightcurve
width-luminosity effect
\begin{equation}
m_B^{\rm corr}  =  m_B  + \alpha (s-1),
\end{equation}
where $s$ is the measured stretch factor and $\alpha$ is a constant to be
determined. These corrected magnitudes are compared to the predicted
magnitudes
\begin{equation}
m_B^{\rm pred} (z) =  {\cal M}_{B} + 5{\rm log}{\cal D}_L(z, \Omega_m, 
\Omega_\Lambda),
\end{equation}
where ${\cal M}_B$ is related to the corrected absoluted
magnitude $M_B$ by ${\cal M}_B = M_B - 5 {\rm log}H_0 + 25$,
and ${\cal D}_L = d_L + 5 {\rm log}H_0$
is the Hubble-constant-free luminosity distance
defined by P98. To compute the luminosity distance, we ignore
gravitational lensing and use the standard expression for
a Universe with uniform density (see {\it e.g.}  Peebles 1993),
%$$
%d_L(z, \Omega_m, \Omega_\Lambda)  =  {c \over H_0} {(1+z) \over \vert
%\Omega_k \vert^{1/2}} \left\{ \begin{array}{l} {\rm sinh} \\ 
%{\rm sin} \end{array}
%\right \} \left [ \vert \Omega_k \vert^{1/2} x(z, \Omega_m,
%\Omega_\Lambda) \right ]
%$$
$$
d_L(z, \Omega_m, \Omega_\Lambda)  =  {c \over H_0} {(1+z) \over \vert
\Omega_k \vert^{1/2}} 
{\rm sin}_k \left [ \vert \Omega_k \vert^{1/2} x(z, \Omega_m,
\Omega_\Lambda) \right ],
$$
$$
x(z, \Omega_m, \Omega_\Lambda) =  
\int_0^z {dz^\prime 
\over 
[\Omega_m (1 + z^\prime)^3 + \Omega_k ( 1 + z^\prime)^2 +
\Omega_\Lambda]^{1/2} }
\quad  (5) 
$$
where $\Omega_k = 1 - \Omega_m - \Omega_\Lambda$ and ${\rm sin}_k =
{\rm sinh}$ if $\Omega_k > 0$ and ${\rm sin_k} = {\rm sin}$ for
$\Omega_k < 0$.  We assume a constant cosmological constant here
rather than an arbitrary equation of state as might arise with scalar
fields that become important at late times (Ratra and Peebles 1988,
Caldwell, Dave and Steinhardt 1998). The biases in the
magnitude-redshift relation arising from gravitational lensing should
be negligible for CDM-like models unless a large fraction of the dark
matter is in compact objects ({\it e.g.} Wambsganss, Cen and Ostriker
1998).  Even in the latter case, P98 show that the biases are
relatively small for the low matter densities favoured by the SN data.
\addtocounter{equation}{1}

With the above assumptions, we perform a four parameter 
($\Omega_m$, $\Omega_\Lambda$,
${\cal M}_B$ and $\alpha$) likelihood
analysis  assuming Gaussian errors 
on $m_B^{corr}$  consisting  of three terms, 
\begin{equation}
(\Delta m_B^{corr})^2 = \Delta m_B^2 + \alpha^2 \Delta s^2
+  \Delta m_{\rm intrinsic}^2, 
\end{equation}
where $\Delta m_B$ and $\Delta s$ are the measurement errors
in $m_B$ and $s$ and $\Delta m_{\rm intrinsic}$ is the instrinsic
dispersion in $m_B$ determined to be $0.18$ mag from the 
Cal\'an/Tololo sample. (The maximimum likelihood parameters are
extremely insensitive to  $\Delta m_{\rm intrinsic}$.)
This analysis differs from that in P98 in that we 
include the dependence of the magnitude errors on the
parameter $\alpha$  self-consistently in the likelihood
analysis via equation (6). As we will see
below this has little effect on the likelihood constraints
on $\Omega_\Lambda$ and $\Omega_m$ (Figure 3), which are
in good agreement with the results presented in P98. However,
our analysis allows an important
test of the intrinsic properties of high and low redshift supernovae
(see Figure 4).

\begin{figure}

\vskip 2.6 truein

\includegraphics{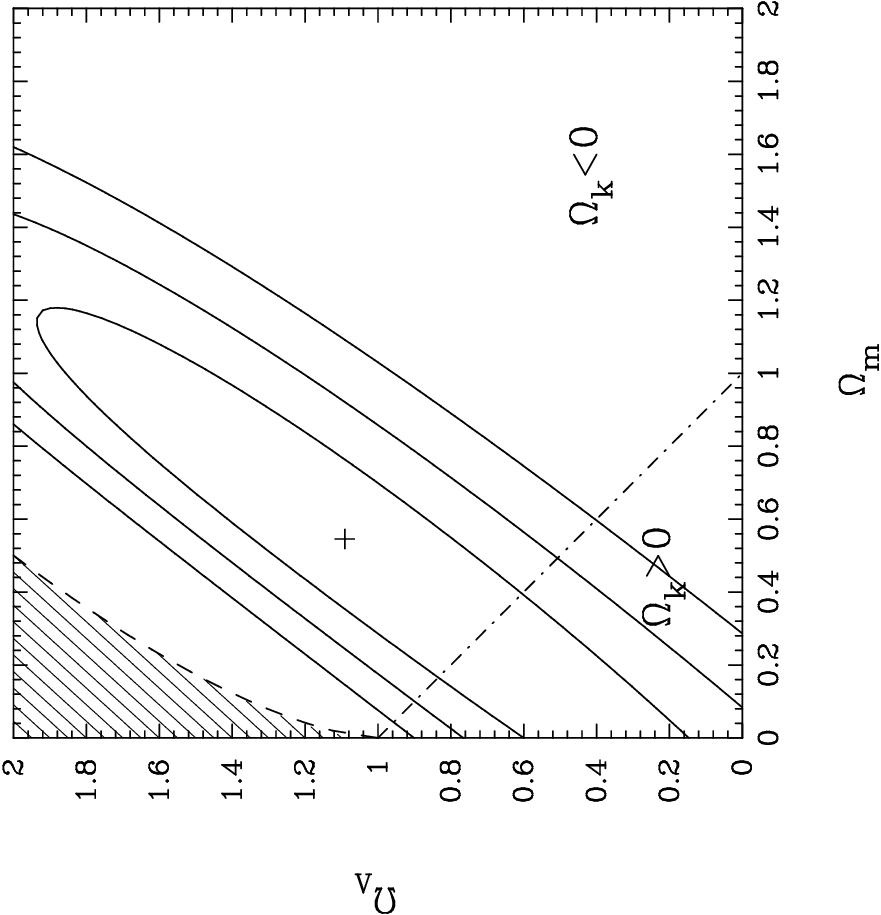}

\caption
{Likelihood contours ($1$, $2$ and $3\sigma$, as defined in the
caption to Figure 2) in the $\Omega_\Lambda$--$\Omega_m$ plane derived
from the SN magnitude-redshift relation after marginalization over the
parameters ${\cal M}_B$ and $\alpha$.
The cross shows the maximum likelihood solution. Bouncing
universes have parameters within  the hatched region.}
\label{figure3}
\end{figure}

Excluding the four SN as described above, we analyse the combined
high redshift P98  + Cal\'an/Tololo samples (denoted `combined'),
and the two samples separately (denoted `high z' and `low z'
respectively). The magnitude-redshift relation of the low z
sample is relatively insensitive to cosmology, hence we fix $\Omega_m =1$ 
and $\Omega_\Lambda =0$ in the likelihood analysis. Table 1 lists the
parameters that maximize the likelihood for these samples.

\begin{table}
\label{tab1}
\centerline{\bf Table 1: Maximum Likelihood Parameters}
\begin{center}
\begin{tabular}{|cccc|l|} \hline
${\cal M}_B$ &  $\alpha$      & $\Omega_m$  & $\Omega_\Lambda$ &  sample  \\ 
$-3.45$        &  $1.33$  &  $0.54$ & $1.09$ & combined \\
$-3.69$        &  $1.41$  &  $0.75$ & $1.86$ & high z \\
$-3.37$        &  $1.32$  &  --    &  --    & low z\\ \hline
\end{tabular} 

\vskip -0.2 truein
\end{center}
\end{table}

Figure 3 shows the likelihood function for the combined sample in the
$\Omega_m$--$\Omega_\Lambda$ plane after marginalization over the
parameters ${\cal M}_B$ and $\alpha$ assuming uniform prior
distributions in these variables. We use this likelihood function in
Section 3 when we combine the SN sample with parameters derived from
the CMB.  As P98 show (and we have confirmed) changes in the analysis,
{\it e.g.} omitting outliers, stretch correction, reddened objects,
usually shifts the error ellipses by much less than the width of the
$1 \sigma$ contour. 

An important consistency check is the agreement between the intrinsic
properties of the high and low redshift SN. This is illustrated in
Figure 4 which shows likelihood contours for the parameters ${\cal
M}_B$ and $\alpha$ for the low z and high z samples (marginalized over
$\Omega_\Lambda$ and $\Omega_m$ for the high z sample). This diagram
shows that the low z and high z samples have the same lightcurve
width-luminosity relation and consistent peak absoluted magnitudes
${\cal M}_B$. (Note, however, that the contours for the high z sample
become highly elongated in the ${\cal M}_B$ direction because this
parameter correlates strongly with $\Omega_\Lambda$ and
$\Omega_m$). We have repeated the likelihood analysis including the
intrinsic magnitude scatter $\Delta m_{\rm intrinsic}$ as a free
parameter. For the low-z sample, the likelihood gives a broad
distribution peaked at $\Delta m_{\rm intrinsic} = 0.18$, consistent
with the {\it rms} residual of $0.16$ magnitude of the Cal\'an/Tololo
points around the best fit solution.  The combined sample gives
$\Delta m_{\rm intrinsic} = 0.20$, slightly higher but consistent
with the low-z sample. (In fact, $\Delta m_{\rm intrinsic}$ drops to
$0.18$ mag if we remove three outliers, SN92bi, SN95as and
SN97K).

\begin{figure}

\vskip 2.2 truein

\includegraphics{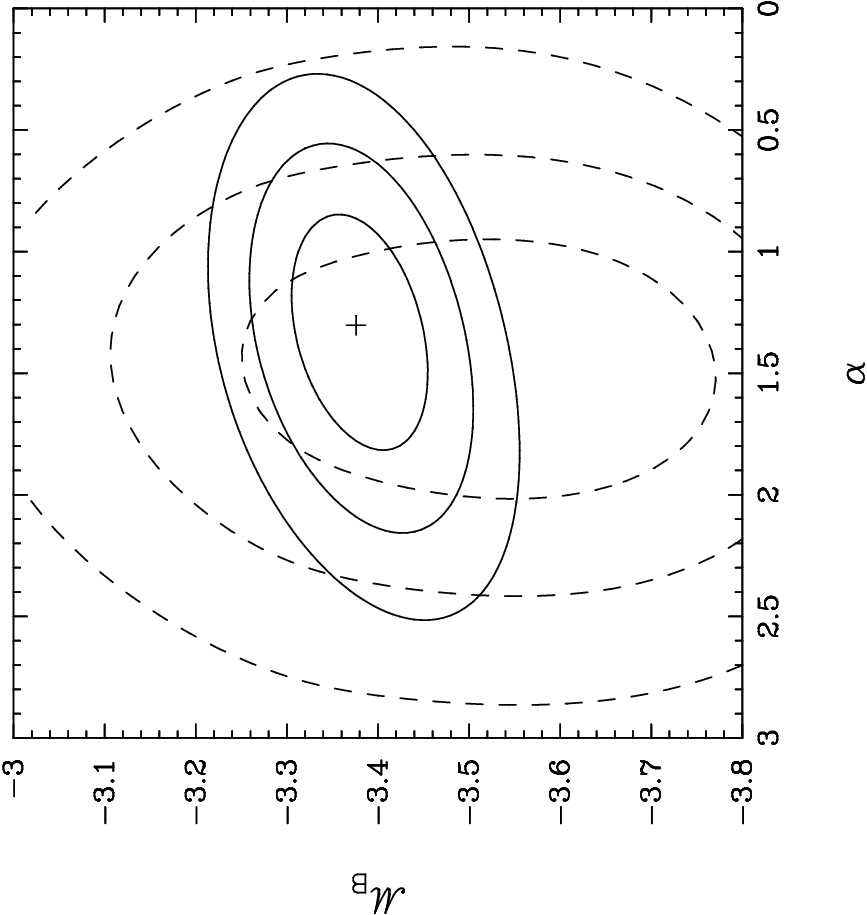}

\caption
{Likelihood contours ($1$, $2$ and $3\sigma$) 
in the ${\cal M}_B$--$\alpha$ plane. The solid lines show 
contours for the Cal\'an/Tololo sample and the dotted lines
show contours for the high redshift P98 sample after marginalization
over the cosmological parameters $\Omega_\Lambda$ and $\Omega_m$.
}
\label{figure4}
\end{figure}

For low redshift supernovae, the lightcurve widths and 
luminosities are known to correlate with the morphological
type of the host galaxies (Hamuy \etal 1996). 
Thus the lack of any significant difference between the high and low redshift
samples in Figure 2 provides a constraint on  evolutionary corrections
associated with systematic changes in the host galaxies with
redshift ({\it e.g.} metallicity, morphological mix). 
Figure 4 therefore indicates no 
detectable evolutionary trends in the underlying supernova population
used in this analysis.

Table 1 shows that the high z sample alone yields 
values for $\Omega_m$ and $\Omega_\Lambda$ that are similar 
to those of the combined sample.
However, the error contours for the high z sample 
are much larger than those shown in Figure 1 for the combined sample
so that an $\Omega_m =1$, $\Omega_\Lambda=0$ universe lies within the
$2\sigma$ contour. The narrowness of the likelihood contours shown in Figure
3 therefore rely on combining the P98 data with the
Cal\'an/Tololo sample, though the general trends are evident in the 
high-z sample alone.

In summary, the P98 and Cal\'an/Tololo SN strongly favour a universe
with $0.78\Omega_m - 0.62 \Omega_\Lambda \approx -0.25 \pm 0.13$.
This is consistent with the analysis of P98 (who find $0.8 \Omega_m -
0.6 \Omega_\Lambda \approx -0.2 \pm 0.1$) and with the analysis of a
sample of $16$ SN at $z>0.16$ ($14$ of which are independent of the
P98 high redshift sample) combined with $34$ low z SN from the
Cal\'an/Tololo and CfA samples (Riess \etal 1998, Garnavich \etal
1998). Furthermore, the lightcurve width-luminosity relation of the
high z SN is consistent with that of the nearby sample suggesting that
these objects have similar intrinsic properties. If an evolutionary
effect is causing a systematic error in the magnitude-redshift
relation of the distant sample, then it must be so as to preserve the
lightcurve width-luminosity relation.

\begin{figure}

\vskip 2.5 truein

\includegraphics{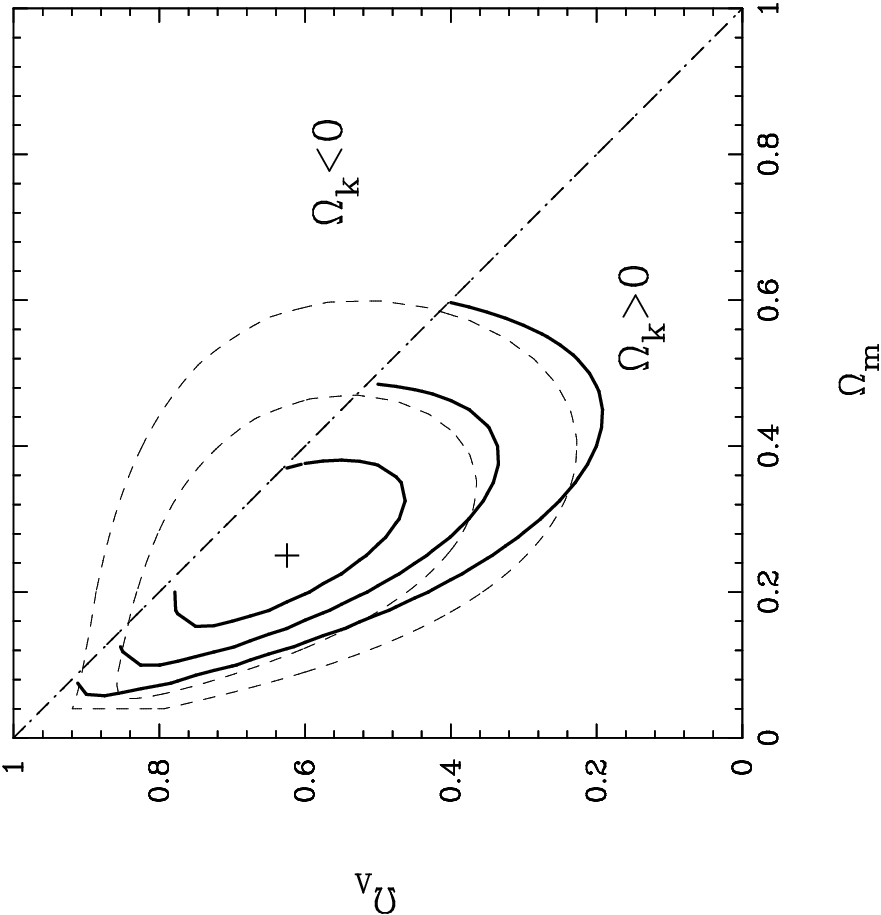}

\caption
{Likelihood  contours derived by combining the
supernovae likelihood function shown in Figure 3 with the
CMB likelihood function shown in Figure 2. As in Figure 2,
solid contours show $1$, $2$ and $3\sigma$
confidence intervals  computed using CMBFAST. The dashed contours
show $2$ and $3 \sigma$ contours computed using the approximate 
CMB fitting technique (for clarity we do not plot the $1\sigma$ contour).
The combined CMB+SN  likelihood function peaks at $\Omega_m=
0.25$  and $\Omega_\Lambda = 0.63$}.
\vskip - 0.2 truein
\label{figure5}
\end{figure}

\section{Combining the Supernovae and  CMB likelihoods}

\subsection{Combined likelihoods}

The combined likelihood obtained by multiplying 
the SN and CMB likelihoods are shown in Figure
5. As in Figure 2, the solid lines show the likelihood computed with
CMBFAST and the dashed lines show the approximate technique indicating
how the contours extend into the $\Omega_k < 0$ regime.  The
likelihood peaks at $\Omega_m = 0.25$ and $\Omega_\Lambda = 0.63$. The combined
likelihood thus strongly favours a nearly spatially flat Universe with
a low matter density and high cosmological constant. In fact, the
$2\sigma$ ellipse in Figure 5 extends over the range $\Omega_m \approx
0.12$, $\Omega_\Lambda = 0.84$, to $\Omega_m \approx 0.49$ and
$\Omega_\Lambda=0.51$. A high value of $\Omega_\Lambda$ is suggested
by the SN data alone, and is required if we impose the constraint
$\Omega_k = 0$ (see Figure 7 of P98; Figure 6 Riess \etal 1998). However,
from the SN data alone we cannot rule out an open Universe with a low
matter density $\Omega_m \simlt 0.1$ and zero cosmological
constant. Since the CMB data favour a universe with $\Omega_k = 0$,
the combined SN+CMB data {\it require a non-zero cosmological constant
at a high significance level}. This is the main result of this paper.

Figure 6 provides another illustration of how the combination of CMB
and SN data dramatically improve the constraints on $\Omega_\Lambda$
and $\Omega_m$. Here we have plotted the likelihood functions
marginalized over all other parameters except $\Omega_m$ (Figure 6a)
and $\Omega_\Lambda$ (Figure 6b) for the SN and CMB data alone and
for the combined data sets. For a Gaussian likelihood function, the
95\% confidence region is delineated by 
${\cal L}/{\cal L}_{\rm max} \ge 0.146$ and so we can see from Figure
6 that the constraints on $\Omega_m$ and $\Omega_\Lambda$ from
the SN and CMB data alone are extremely weak. However, for the
combined data sets we find $\Omega_m = 0.25^{+0.18}_{-0.12}$ and
$\Omega_\Lambda = 0.63^{+0.17}_{-0.23}$ at the 95\% confidence
level.

\begin{figure}

\vskip 2.5 truein

\includegraphics{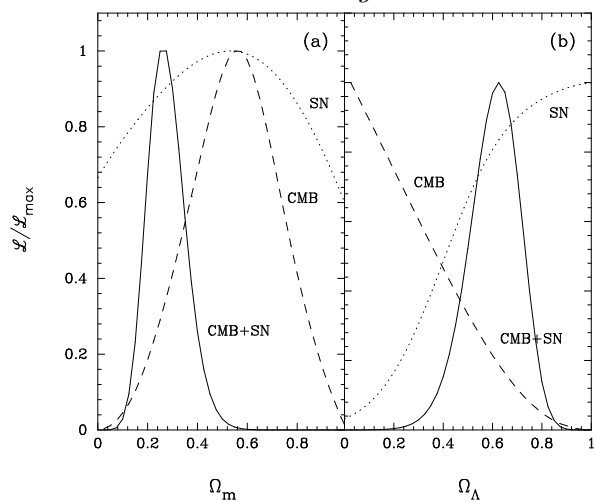}

\caption
{Marginalized likelihood functions plotted as function of $\Omega_m$ (left hand
panel) and $\Omega_\Lambda$ (right hand panel). The dotted lines are
for the SN data alone (Figure 3), dashed lines for the CMB data
alone (Figure 2) and solid lines for the combined data-sets 
(Figure 5).}
\label{figure6}
\end{figure}  

\vskip -0.15 truein

\subsection{Systematic errors}

Systematic errors in the analysis of the SN data are discussed at
length by P98 and by Riess \etal 1998. No systematic error has yet
been identified that could produce a large downward shift of the SN
likelihood contours in Figure 3. Possible sources of systematic error
include differences in the reddening caused by the host galaxies, and
evolutionary ({\it e.g.} metallicity dependent) corrections to the SN
absolute magnitude-redshift relation. Internal reddening for this
sample is discussed in detail by P98. They find no excess reddening
of most of the distant SN when compared to the Cal\'an/Tololo sample,
although there are a small number of possibly reddened SN (three of
which have been excluded in this analysis).  The inclusion or
exclusion of these reddened objects does not significantly affect
Figure 5. Grey extinction is much harder to rule out, but may not
be physically well motivated.

Possible evolutionary effects are difficult to check. However, the
results summarized in Figure 4 show that the high and low redshift SN
sample have statistically indistinguishable internal properties, {\it
i.e.} they have consistent lightcurve width-luminosity relations and
(within rather large errors) consistent peak absolute magnitudes
${\cal M}_B$. With a larger sample of SN it should be possible to
refine this test and to test for differences in the distribution of
lightcurve shapes with redshift. Another consistency check would be
provided by extending the SN sample to $z > 1$.  As Goobar and
Perlmutter (1995) have discussed, the degeneracy of the
magnitude-redshift relation in the $\Omega_m-\Omega_\Lambda$ plane can
be broken by a sample of SN spanning a sufficiently wide range of
redshifts.

\subsection{The best fit Universe}

 If systematic errors are indeed small, the combined CMB and SN data
strongly favour a near-spatially flat Universe with $\Omega_m \approx
0.25$ and $\Omega_\Lambda \approx 0.63$.  These values are close to those
favoured by a number of other arguments, which we summarize briefly below
(see also {\it e.g.} Ostriker and Steinhardt 1995, P98, and references
therein).

\noindent
{\it Age \& Hubble constant:} For our best fit cosmology,
the age of the Universe is $14.6 (h/0.65)^{-1}$ Gyr (in agreement with
P97). This is
compatible with recent estimates of $11.5 \pm 1.3$ Gyr
for the ages of the oldest globular clusters (see Chaboyer 1998) 
and with recent values of $H_0$ derived from Type 1a
supernovae and Cepheid distances,  which  fall  within the 
range $H_0 = 65 \pm 10\; {\rm km} {\rm s}^{-1} {\rm Mpc}^{-1}$
({\it e.g.} Freedman \etal 1998).

\noindent
{\it Large-scale structure:} Observations
of large-scale structure (see {\it e.g.} Efstathiou 1996
for a review) are consistent with scale-invariant adiabatic
cold dark matter universes if $\Gamma \approx \Omega_m h \approx
0.2$--$0.3$. This is broadly consistent with the analysis presented
here and with the analysis of combined CMB and IRAS galaxy data
presented by Webster \etal (1998).

\noindent
{\it Baryon abundance in clusters:} Consistency between primordial 
nucleosynthesis ($\omega_b \approx 0.019$) and the ratio of baryons
in clusters to total cluster mass ($f_b \approx 0.06 h^{-3/2}$,
see Evrard 1998 and references therein) requires a low matter
density, $\Omega_m \approx 0.26 (h/0.65)^{-1/2}$, consistent
with the best fit solution of Figure 5.

\section{Conclusions}

\noindent
$\bullet$ We have applied an approximate formula for the CMB power
spectrum that can be used to constrain a wide set of cosmological
parameters, including closed universes, by fitting to the CMB
anisotropy data.  The results agree well with those derived from
exact computations using the CMBFAST code.

\noindent
$\bullet$ In our analysis we perform a proper marginalization over
parameters, assuming uniform prior distributions,
to derive constraints in the $\Omega_m$-$\Omega_\Lambda$ plane and on
$\Omega_m$ and $\Omega_\Lambda$ separately.

\noindent
$\bullet$ Current CMB anisotropy data provide strong constraints on
the position of the first Doppler peak and favour a spatially flat 
Universe.

\noindent
$\bullet$ A likelihood analysis of the SN data provides robust 
constraints on $\Omega_m$ and $\Omega_\Lambda$ consistent with those
derived by P98. For a spatially flat Universe, the SN data require
a non-zero cosmological constant at a high level of
significance ($\Omega_\Lambda \simgt 0.5$ at 95 \% confidence).

\noindent
$\bullet$ The lightcurve width-luminosity relation for high redshift
and low redshift SN are statistically indistinguishable, consistent
with no  evolution of the SN population.

\noindent
$\bullet$ The combination of CMB and SN data thus provide strong
constraints on $\Omega_m$ and $\Omega_\Lambda$ favouring values of
$\Omega_m \approx 0.25$ and $\Omega_\Lambda \approx 0.63$. If there
are no significant systematic effects in the SN
data ({\it e.g.} grey dust, evolution) then we are forced into
accepting a cosmological constant or a ``quintessence''-like component
of the Universe (Caldwell \etal 1998, Garnavich \etal 1998). 

\noindent
$\bullet$ There are a number of independent lines of argument, {\it
e.g.} the age of the Universe, large-scale clustering of galaxies and
the baryon content of clusters, to support
the best-fit parameters derived in this paper. 

\vskip 0.2 truein 

\noindent
{\bf Acknowledgements.} G. Efstathiou thanks PPARC for the award of a
Senior Fellowship.  S. Bridle acknowledges the receipt of a PPARC
studentship. We thank the members of the Supernova Cosmology Project
for allowing us to use their data, especially Saul Perlmutter, who
provided many helpful comments on this manuscript, and also Mike Irwin
and Gerson Goldhaber for discussions of the P98 analysis.


\begin{thebibliography}{}

%\bibitem[Baker \etal 1998]{cat2} Baker J.C. \etal, 1998, MNRAS, submitted

%\bibitem[Bennett \etal 1996]{bennett} Bennett C.L. \etal, 1996, ApJ., 464, L1

\bibitem[\protect\citename{Bond }1996]{B96} 
Bond, J.R. 1996, {\it Theory and Observations of the
Cosmic Background Radiation}, in ``Cosmology and Large Scale
Structure'', Les Houches Session LX, August 1993, 
eds. R. Schaeffer, J. Silk, M. Spiro and J. Zinn-Justin, 
Elsevier Science Press, Amsterdam,  p469.

\bibitem[\protect\citename{BET}1997]{BET97}
Bond J.R., Efstathiou G., Tegmark M., 1997, MNRAS, 291 L33.

\bibitem[\protect\citename{BondJaffe }1997]{bj97} 
Bond, J.R., Jaffe, A. 1997, in Proc. XXXI Rencontre de Moriond, 
ed. F. Bouchet, Edition Fronti\`{e}res, in press; astro-ph/9610091. 

\bibitem[Bond 1998]{BKN98} 
Bond J.R., Jaffe A.H. and Knox L.E.,  1998.  astro-ph/9808264.

\bibitem[\protect\citename{Burlesa}1998a]{BT98a}
Burles S., Tytler D.,  1998a,   to appear in the 
Proceedings of the Second Oak Ridge
 Symposium on Atomic \& Nuclear Astrophysics, 
ed.   A. Mezzacappa, Institute of Physics, Bristol. astro-ph/9803071

\bibitem[\protect\citename{Burlesb}1998b]{BT98b}
Burles S., Tytler D.,  1998b,  ApJ, in press. astro-ph/9712109

\bibitem[\protect\citename{Caldwell98}1998]{quintessence}
Caldwell, R.R., Dave, R., Steinhardt  P.J., 1998, Phys Rev Lett, 80, 1582.

%\bibitem[\protect\citename{CCM}1989]{CCM89}
%Cardelli J.A., Clayton G.C., Mathis J.S.,   1989, ApJ, 345, 245.

\bibitem[\protect\citename{CPT}1992]{CPT92}
Carroll S.M., Press W.H., Turner E.L., 1992, Ann. Rev. Astr. 
Astrophys., 30, 499.

\bibitem[\protect\citename{Chaboyer}1998]{C98} 
Chaboyer B., 1998, astro-ph/9808200.

%\bibitem[Cheng \etal 1994]{msam1} 
%Cheng E.S. \etal, 1994, ApJ., 422, L37.
%\bibitem[Cheng \etal 1996]{msam2} 
%Cheng E.S. \etal, 1996, ApJ., 456, L71.


\bibitem[\protect\citename{Devlin}1998]{D98} 
Devlin M.J., De Oliveira-Costa A., Herbig T., Miller A.D., Netterfield C.B.,
Page L., Tegmark M., 1998, submitted to ApJL. astro-ph/9808043.

\bibitem[\protect\citename{Efstathiou}1996]{E96} 
Efstathiou G. 1996, {\it Observations of Large-Scale Structure in the
Universe}, in ``Cosmology and Large Scale
Structure'', Les Houches Session LX, August 1993, eds. R. Schaeffer,
J. Silk, M. Spiro and J. Zinn-Justin, 
Elsevier Science Press, Amsterdam,  p135.

\bibitem[\protect\citename{EB}1998]{EB98}
Efstathiou G.,  Bond J.R., MNRAS in press. astro-ph/9807130.

\bibitem[\protect\citename{EV}1998]{EV98}
Evrard G., 1998, submitted to MNRAS. astro-ph/9701148.

\bibitem[\protect\citename{Freedman}1998]{FMKM98} 
Freedman J.B., Mould J.R., Kennicutt R.C., Madore B.F., 1998, 
astro-ph/9801090.

\bibitem[\protect\citename{Garnavich}1998]{Garnavich98}
Garnavich P.M. \etal 1998,  astro-ph/9806396.

\bibitem[\protect\citename{Goobar}1995]{GP95}
Goobar A., Perlmutter S., 1995, ApJ, {450}, 14.

%\bibitem[Gundersen \etal 1995]{spole} 
%Gundersen J.O \etal, 1995, ApJ., 443, L57

%\bibitem[Gutierrez 1997]{guti97} 
%Gutierrez C.M., 1997, ApJ., 483, 51

\bibitem[\protect\citename{Hamuy}1996]{H96}
Hamuy M., Phillips M.M., Maza J., Suntzeff N.B., Schommer R.A.,
Aviles R. 1996, AJ, 112, 2391.

%\bibitem[Hancock \etal 1994]{nature94} 
%	Hancock S., \etal, 1994, Nature, 367, 333

\bibitem[Hancock \etal 1997]{me96} 
Hancock S., Gutierrez C.M., Davies R.D., Lasenby A.N., Rocha G.,
Rebolo R., Watson R.A., Tegmark M., 1997, MNRAS, 298, 505

\bibitem[Hancock \etal 1998]{sh97}
	Hancock S., Rocha G., Lasenby A.N., Gutierrez C.M., 1998,
        MNRAS, 294, L1

\bibitem[\protect\citename{Herbig}1998]{H98} 
Herbig T.,  De Oliveira-Costa A., Devlin  M.J.,  Miller A.D.,
Page L.,  Tegmark M., 1998, submitted to ApJL. astro-ph/9808044.

%\bibitem[Leitch \etal 1998]{ovro}
%Leitch E.M., Readhead A.C.S., Pearson T.J., Myers S.T., \& Gulkis
%S., 1998, astro-ph/9807312

\bibitem[Lineweaver 1998]{L98}
Lineweaver C.H., 1998. ApJ, {505}, L69.

\bibitem[\protect\citename{Lineweaver98a}1998]{LB98a}
Lineweaver, C.H., Barbosa D., 1998a, ApJ, {446}, 624.

\bibitem[\protect\citename{Lineweaver98b}1998]{LB98b}
Lineweaver, C.H., Barbosa D., 1998b, A\&A, 329, 799.

%\bibitem[Netterfield \etal 1997]{sask} 
%Netterfield C.B., Devlin M.J., Jarosik N., Page L., 
%       Wollack E.J., 1997, ApJ, 474, 47

\bibitem[\protect\citename{Oliveira-Costa}1998]{OC98} 
De Oliveira-Costa A., Devlin  M.J.,  Herbig T., Miller A.D.,
Netterfield C.B. Page L.,  Tegmark M., 1998, submitted to ApJL. 
astro-ph/9808045.

\bibitem[\protect\citename{Ostriker}1995]{OS95}
Ostriker J.P., Steinhardt P.J.,  1995, Nature, {377}, 600.


\bibitem[\protect\citename{Peebles93}1993]{P93}
Peebles P.J.E., 1993, {\it Principles of Physical Cosmology}, 
Princeton University Press, Princeton, New Jersey.

\bibitem[\protect\citename{Perlmutter95}1995]{P95}
Perlmutter S, et al., 1995, In Presentations at the NATO ASI in
Aiguablava, Spain, LBL-38400; also published in Thermonuclear
Supernova, P. Ruiz-Lapuente, R. Cana and J. Isern (eds),
Dordrecht, Kluwer, 1997, p749.

\bibitem[\protect\citename{Perlmutter97a}1997]{P97}
Perlmutter S, et al., 1997, ApJ, 483, 565.

\bibitem[\protect\citename{Petalb}1998]{P98a}
Perlmutter S. \etal, 1998a, ApJ, in press. (P98). astro-ph/9812133.

\bibitem[\protect\citename{Petala}1998]{P98a}
Perlmutter S. \etal, 1998b, In Presentation at the January 1988
Meeting of the American Astronomical Society, Washington D.C.,
LBL-42230,
available at www-supernova.lbl.gov; B.A.A.S., volume 29, p1351, 1997.



\bibitem[\protect\citename{Petal8c}1998]{P98c}
Perlmutter S, et al., 1998c, Nature, 391, 51.

%\bibitem[Platt \etal 1997]{platt}
%Platt S.R., Kovac J., Dragovan M., Peterson J.B., \& Ruhl J.E., 
%	1997, ApJ, 475, L1

\bibitem[\protect\citename{Ratra88}1988]{latetimescalar}
Ratra B., Peebles P.J.E., 1988, Phys Rev D 37, 3406.

\bibitem[\protect\citename{Riess}1998]{Riess98}
Riess A. \etal 1998, AJ, in press. astro-ph/9805201.

%\bibitem[Rocha \etal 1998]{gr}
%Rocha, G., Hancock, S., Lasenby, A.N., \& Gutierrez, C.M. 
%        in preparation.

%\bibitem[Ruhl \etal 1995]{python} 
%Ruhl J.E., Dragovan M., Platt S.R., Kovac J., \& Novak G.,
%	1995, ApJ., 453, L1

%\bibitem[\protect\citename{SFD}1998]{SFD98}
%Schlegal D., Finkbeineer D., Davis M.,  1998, ApJ in press. 
%astro-phxxxx.

%\bibitem[Scott \etal 1996]{cat} 
%	Scott, P.F. \etal, 1996, ApJ., 461, L1

\bibitem[\protect\citename
{Seljak \& Zaldarriaga }1996]{SZ96}
Seljak U.,  Zaldarriaga M. 1996, ApJ,  469, 437.


\bibitem[\protect\citename{SZ}1998]{SZ98}
Seljak U. \& Zaldarriaga M., 1998,  astro-ph/9811123.

%\bibitem[Smoot \etal 1992]{smoot} 
%Smoot G.F. \etal, 1992, ApJ., 396, L1





%\bibitem[Tanaka \etal 1996]{max} 
%Tanaka, S.T. \etal, ApJ, 468, L81

\bibitem[\protect\citename
{Tegmark97}1997]{T97}
Tegmark M., 1997, Phys Rev. Lett, 79, 3806.

\bibitem[\protect\citename{Tegmark}1998]{T98} 
Tegmark M. 1998, submitted to  ApJL. 
astro-ph/9809201.

\bibitem[\protect\citename{Tegmarketal98}1998]{TEHK98} 
Tegmark, M., Eisenstein D.J., Hu W., Kron R.G., 1998, astro-ph/9805117.

\bibitem[\protect\citename{WCO}1998]{WCO98}
Wambsganss J., Cen R., Ostriker J.P., 1998, 
ApJ, 494, 29.

\bibitem[\protect\citename{Webster98}1998]{Web98}
Webster M., Bridle S.L., Hobson M.P., Lasenby A.N., Lahav O., Rocha, G., 1998,
ApJL, in press, 
astro-ph/9802109.


\bibitem[\protect\citename{White98}1998]{W98}
White M., 1998, ApJ in press. astro-ph/9802295.

\bibitem[\protect\citename{Zaldarriaga97}1997]{ZSS97}
Zaldarriaga, M., Spergel D.N., Seljak U.,
1997 ApJ, 488, 1.

\end{thebibliography}
\end{document}